\documentclass[10pt]{article}
\usepackage[utf8]{inputenc}
\usepackage{amsmath, graphicx}
\usepackage[a4paper]{geometry}

\begin{document}
\title{Nanobubble-induced flow of immersed glassy polymer films}
\author{Christian Pedersen$^1$ \and Shuai Ren$^2$ \and Yuliang Wang$^{2, 3}$ \and Andreas Carlson$^1$ \and Thomas Salez$^{4,5,*}$}

\date{%
\scriptsize $^1$Mechanics Division, Department of Mathematics, University of Oslo, 0316 Oslo, Norway.\\
$^2$School of Mechanical Engineering and Automation, Beihang University, 37 Xueyuan Rd., Haidian District, Beijing 100191, China.\\
$^3$Beijing Advanced Innovation Center for Biomedical Engineering, Beihang University, 37 Xueyuan Rd., Haidian District, Beijing 100191, China.\\
$^4$Univ. Bordeaux, CNRS, LOMA, UMR 5798, F-33405 Talence, France.\\
$^5$Global Station for Soft Matter, Global Institution for Collaborative Research and Education, Hokkaido University, Sapporo, Hokkaido 060-0808, Japan.\\
$^*$thomas.salez@u-bordeaux.fr
}

\maketitle

\begin{abstract}
\noindent We study the free-surface deformation dynamics of an immersed glassy thin polymer film supported on a substrate, induced by an air nanobubble at the free surface.
We combine analytical and numerical treatments of the glassy thin film equation, resulting from the lubrication approximation applied to the surface mobile layer of the glassy film, under the driving of an axisymmetric step function in the pressure term accounting for the nanobubble's Laplace pressure.
 Using the method of Green's functions, we derive a general solution for the film profile. We show that the lateral extent of the surface perturbation follows an asymptotic viscocapillary power-law behaviour in time, and that the film's central height decays logarithmically in time in this regime. This process eventually leads to film rupture and dewetting at finite time, for which we provide an analytical prediction exhibiting explicitly the dependencies in surface mobility, film thickness and bubble size, among others. Finally, using finite-element numerical integration, we discuss how non-linear effects induced by the curvature and film profile can affect the evolution.
\end{abstract}

\section*{Introduction}

The formation and rheological properties of glassy materials have been of great interest to the scientific community for many decades~\cite{berthier2011theoretical}.
The relation between the viscosity of glass-forming materials and temperature can be divided into two main trends: the so-called strong and fragile supercooled liquids~\cite{martinez2001thermodynamic}. The former category exhibits an Arrhenius-like temperature dependence of the viscosity, reminiscent of the behaviour of simple liquids, whereas the latter category exhibits an apparent divergence of the viscosity at finite temperature. This suggests the existence of two very different relaxation processes at the molecular level~\cite{kivelson2008search} -- the dynamics of fragile supercooled liquids being often associated to a cooperative relaxation process, and thus to the existence of some characteristic supermolecular sizes.

The possible existence of such a cooperative length scale has led to an intense research activity around confined glass formers, with perhaps the most emblematic role played by thin polymer films~\cite{forrest2001glass,forrest2002decade,ediger2012perspective}. In such samples, it has been reported that the glass transition temperature $T_{\textrm{g}}$ is typically lower than for their bulk counterparts, and heavily influenced by the film thickness~\cite{PhysRevLett.77.2002, peter2006thickness}. These observations were further connected with the dynamics at the free surface~\cite{PhysRevLett.109.055701}. Therein, a liquid-like surface layer of nanometric thickness  was reported~\cite{ilton2009using, yang2010glass, kim2018direct}, with no dependency on film thickness or molecular weight~\cite{paeng2011direct}.
In a similar fashion as the surface diffusion of crystals, the enhanced surface mobility thus also appears as a characteristic feature of amorphous solids, allowing for surface tension to smoothen out asperities~\cite{bartak2021surface,hornat2020entropy}. One seminal method to study this enhanced surface mobility is by embedding small gold particles into the glass surface~\cite{ilton2009using,qi2011measuring}. The embedded particles exhibit a decreasing degree of mobility with increasing particle size and embedding depth, highlighting the finite thickness of the mobile layer.  Another method consists in studying the levelling dynamics of free-surface perturbations. For polymers above $T_{\textrm{g}}$, this has been a successful method to characterize bulk dynamics~\cite{cormier2012beyond,benzaquen2014approach,bertin2021capillary}. In such melt conditions, the viscosity is assumed to be homogenous, and the interfacial dynamics can be described by the capillary-driven thin film equation ~\cite{blossey2012thin}.
In contrast, below $T_{\textrm{g}}$, the flow becomes localized within the thin surface mobile layer, making the predictions from the previous bulk equation incorrect. Assuming at lowest order a two-layer model, with i) a mobile layer of nanometric thickness and finite viscosity, atop ii) an immobile bulk layer of infinite viscosity (although the viscosity increases continuously towards the bulk from the free surface ~\cite{forrest2013can}), a glassy thin film model was derived~\cite{salez2012capillary} and shown to be in excellent agreement with experimental observations~\cite{chai2014direct,chai2020using}. While obtained from lubrication theory, the underlying partial differential equation is in fact mathematically equivalent to the one describing surface diffusion, the physical connection between the two descriptions relying on Stokes-Einstein arguments, and thus equilibrium properties of the surface layer. Finally, the solution of the glassy thin film equation is controlled by a single parameter which, after fitting to experimental data, characterizes the mobility of the surface layer below $T_{\textrm{g}}$~\cite{yu2016surface, zhang2017decoupling, chai2020using}.

In contrast to the passive levelling protocole above, recent experiments have demonstrated that immersing a polystyrene film into a bath of water leads to the spontaneous nucleation of air nanobubbles at its free surface~\cite{wang2008nanoindents, RevModPhys.87.981, tarabkova2016single}, a phenomenon attributed to surface roughness~\cite{li2014micro}.
Due the large curvature of such nanobubbles, the internal Young-Laplace pressure can reach several atmospheres. Therefore, they can drive the underlying polymer mobile layer to flow, dynamically deforming the free surface and creating a growing nanocrater underneath.
Beyond potential strategies towards the spontaneous fabrication of smart patterns and porous membranes, this process can be used as a simple and efficient tool to probe the fundamental rheology of glassy surfaces at room temperature. In a previous work~\cite{ren2020capillary}, we demonstrated the robustness of numerical treatments of the glassy thin film equation in 3D with a driving pressure source,  in quantitatively rationalizing atomic force microscope experiments on the nanobubble-induced formation and evolution of surface nanocraters.
While thin film equations with external pressure terms have already been studied~\cite{alleborn2004local,PhysRevFluids.3.114801,ledesma2016wake}, the case of nanobubble-induced glassy surface flows is still open for analytical investigations.

In this article, we combine analytical and numerical treatments of the glassy thin film equation to provide new insights into this problem. The Young-Laplace pressure of the nanobubble is modelled as an axisymmetric step function in the governing equation. We use the method of Green's functions to derive a general solution for the film profile, from which we extract important physical parameters,
such as the central depth, the half width and the excess surface energy. Finally, we investigate how the dynamics is affected when non-linear curvature effects come into play, or when the film thickness becomes similar to that of the mobile layer itself. These can have a large impact, depending on the nanobubble size and film thickness.

\section*{Physical model}

\begin{figure}
\centering
\includegraphics[scale=0.7]{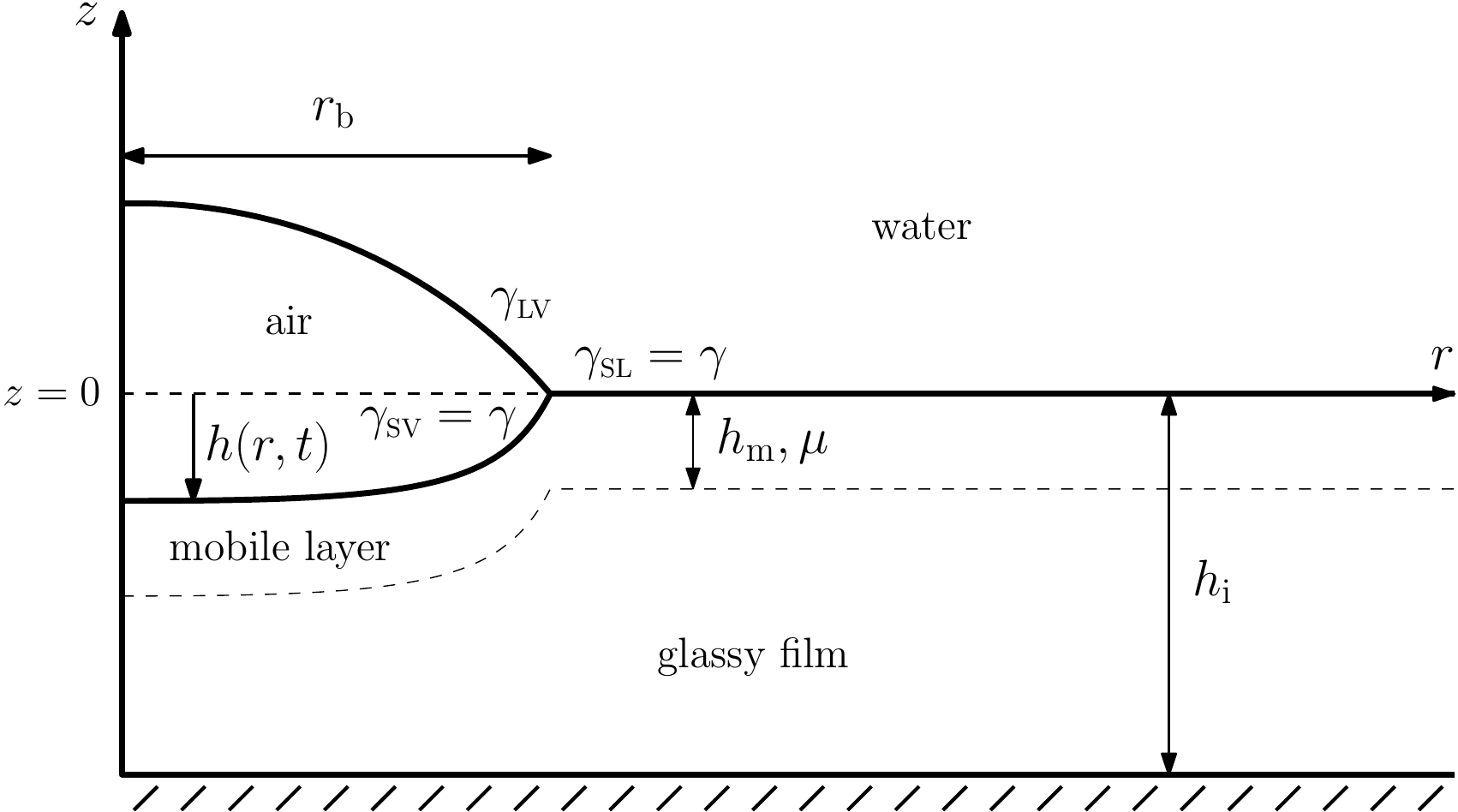}
\caption{Schematic of the system (not drawn to scale). A thin glassy polymer film of initial thickness $h_{\textrm{i}}$, with a thin mobile surface layer of thickness $h_{\textrm{m}}$ and viscosity $\mu$, is immersed in water. At time $t=0$, an axisymmetric air bubble of radius $r_{\textrm{b}}$, with origin at $r=0$, is placed atop the glassy film. The excess Laplace pressure inside the bubble drives the mobile layer to flow and deforms the glassy film with vertical deflection $h(r,t)$. $\gamma_{\textrm{LV}}$, $\gamma_{\textrm{SV}}$ and $\gamma_{\textrm{SL}}$ are the surface tensions of the water-air, film-air and film-water interfaces, respectively. In the mathematical model, we further assume for simplicity that $\gamma_{\textrm{SL}}=\gamma_{\textrm{SV}}$ (further noted $\gamma$), with no loss of generality.}
\label{fig:schematic}
\end{figure}

We consider an initially-flat thin film of thickness $h_{\textrm{i}}$, immersed in a bath of water, see Fig.~\ref{fig:schematic}. The film is below the glass transition temperature and is therefore assumed to be
solid in the bulk, but to have a thin incompressible mobile layer of thickness $h_{\textrm{m}}$ and viscosity $\mu$ at the free surface. The latter is then driven to flow at time $t=0$ by the presence of a surface air bubble. We note $h(r,t)$ the vertical deflection of the interface, in axisymmetric coordinates. By considering nanofilms made of polystyrene at ambient conditions~\cite{ren2020capillary}, we have the following estimates for the surface layer mobility $h_{\textrm{m}}^3/3\mu\sim 10^{-10}\,\textrm{nm}^3\textrm{Pa}^{-1}\textrm{s}^{-1}$, the thickness $h_{\textrm{m}}\sim 10$~nm of the mobile layer,
the film surface tension $\gamma \sim 50\, \textrm{mN/m}$ and the film density $\bar \rho \sim 10^3\, \textrm{kg}/\textrm{m}^3$. From these material parameters, we get the viscosity $\mu\sim10^{13}\, \textrm{Pa s}$, the capillary velocity $v_{\textrm{c}}=\gamma/\mu \sim 1\,\textrm{n\AA}/\textrm{s}$, the Reynolds number $\textrm{Re}=h_{\textrm{m}}\rho v_{\textrm{c}}/\mu \ll 1$ and the capillary length $L_{\textrm{c}} = [\gamma / (\bar\rho\textrm{g})]^{1/2} \sim \, 2\, \textrm{mm}$.
We are thus in a viscous regime, and we can neglect effects from gravity. Moreover, we assume $h_{\textrm{m}}\ll r_{\textrm{b}}$ and employ lubrication theory~\cite{batchelor2000introduction}. This implies  that the vertical pressure gradient is negligible compared to the radial one.
When applying a no-shear boundary condition at the film's free surface $z=h(r,t)$ and a no-slip boundary condition where the mobile layer meets the glassy bulk region, i.e. $z=h(r,t)-h_{\textrm{m}}$,
we obtain the radial velocity component in the mobile layer

\begin{equation}
u(r,t) = \frac{1}{2\mu}\left(z^2 + h^2 - h_{\textrm{m}}^2 - 2hz\right)\frac{\partial p(r,t)}{\partial r}
\label{eq:velocity}
\end{equation}
with $p(r,t)$ the pressure field in the mobile layer. By imposing volume conservation \cite{batchelor2000introduction} we get

\begin{equation}
\frac{\partial h}{\partial t} = \frac{h_{\textrm{m}}^3}{3\mu r}\frac{\partial}{\partial r}\left(r\frac{\partial p}{\partial r}\right).
\label{eq:general_gtfe}
\end{equation}

At time $t=0$, an axisymmetric air nanobubble is placed on top of the film with its center at $r=0$.
Due to the large curvature of the bubble, the Young-Laplace pressure will force a radial flow within the mobile layer and deform the surface. This deformation is in turn opposed by the surface tension
force. For simplicity, we assume that the surface tensions at the film-air and film-water interfaces are
both equal, and we note them $\gamma$. At all times, one has

\begin{equation}
\begin{split}
p - p_0 &= -\gamma\nabla_r\cdot\frac{\mathbf{n}}{|\mathbf{n}|} + p_{\textrm{b}}(r)\Theta(t)\\
        &\simeq -\gamma\frac{1}{r}\frac{\partial}{\partial r}\left( r\frac{\partial h}{\partial r}\right) + p_{\textrm{b}}(r)\Theta(t)
\end{split}
\label{eq:pressure}
\end{equation}
where $\Theta$ is the Heaviside function, $\nabla_r$ is the nabla operator in cylindrical coordinates, $p_0$ is the ambient pressure, $p_{\textrm{b}}(r)=(2\gamma_{\textrm{LV}}/r_{\textrm{b}})\Theta(r_{\textrm{b}}-r)$ is the excess bubble pressure field, and $\mathbf{n}$ is the surface normal vector with norm $|\mathbf{n}|=\sqrt{1+(\partial h/\partial r)^2} \simeq 1$, assuming small slopes. Inserting this expression into Eq.~\eqref{eq:general_gtfe}, we get the axisymmetric glassy thin film equation with external driving pressure:

\begin{equation}
\begin{split}
\frac{\partial h(r,t)}{\partial t} + \frac{h_{\textrm{m}}^{\,3}}{3\eta r} \frac{\partial}{\partial r}\left\{r\frac{\partial}{\partial r}\left[\frac{\gamma}{r}\frac{\partial}{\partial r}\left(r\frac{\partial}{\partial r} h(r,t)\right) - p_{\textrm{b}}(r)\Theta(t) \right] \right\} = 0\ .
\end{split}
\label{eq:gtfe}
\end{equation}
Note that the constant $p_0$ disappears due to the radial spatial derivative.
Finally, we nondimensionalize Eq.~\eqref{eq:gtfe} by introducing $h=h_{\textrm{m}} H$, $r=r_{\textrm{b}} R$  and $t=\tau T$, where $\tau = 3\mu r_{\textrm{b}}^4 / (\gamma h_{\textrm{m}}^3)$ is the characteristic time scale
of the viscocapillary response. Doing so, we obtain the dimensionless form

\begin{equation}
\frac{\partial H(R,T)}{\partial T} + \frac{1}{R} \frac{\partial}{\partial R}\left\{R\frac{\partial}{\partial R}\left[\frac{1}{R}\frac{\partial}{\partial R}\left(R\frac{\partial}{\partial R} H(R,T)\right) - \beta\Theta(T)\Theta(1-R) \right] \right\} = 0
\label{eq:gtfeadim}
\end{equation}
where $\beta = 2r_{\textrm{b}}\gamma_{\textrm{LV}}/(\gamma h_{\textrm{m}})$ is the dimensionless bubble pressure magnitude. To facilitate the understanding of the procedure in the following section, we also include a version of Eq.~\eqref{eq:gtfeadim} in Cartesian coordinates

\begin{equation}
\frac{\partial H(X,Y,T)}{\partial T} + \nabla^2\left[\nabla^2 H(X,Y,T) -  \beta\Theta(T)\Theta\left(1-\sqrt{X^2+Y^2}\right) \right] = 0\
\label{eq:gtfeadim_carth}
\end{equation}
where $\nabla$ is the nabla operator in Cartesian coordinates.

\section*{Green's function}
The Green's function is defined to be the solution of the equation
\begin{equation}
\mathcal{L}G(X,Y,T) = \delta(X,Y,T)
\label{eq:greens_definition}
\end{equation}
where $\mathcal{L} = \partial_T + (\partial_X^2 + \partial_Y^2)^2$ is the linear differential operator of Eq.~\eqref{eq:gtfeadim_carth} and $\delta(X,Y,T)$ is the Dirac delta function. From the Green's function, we can then later obtain the analytical solution for the film thickness profile by solving the convolution
\begin{equation}
H(X,Y,T) = \int \textrm{d}X'\textrm{d}Y'\textrm{d}T'\, G(X-X', Y-Y',T-T') \nabla^2 P_{\textrm{b}}(X',Y',T'),
\label{eq:convolution}
\end{equation}
with $P_{\textrm{b}}(X,Y,T)= \beta\Theta(T)\Theta\left(1-\sqrt{X^2+Y^2}\right)$.

To obtain the Green's function we invoke the Fourier transform
\begin{equation}
\hat{G}(k_X,k_Y,\omega) = \int \textrm{d}X\, \textrm{d}Y \textrm{d}T\, G(X,Y,T) \,\textrm{e}^{-i(k_XX + k_YY + \omega T)}
\end{equation}
where $k_X, k_Y$ are the spatial angular wavenumbers in the $X, Y$ directions, respectively, and $\omega$ is the angular frequency.
When applied to Eq.~\eqref{eq:greens_definition}, we find

\begin{equation}
\hat{G}(k_X,k_Y,\omega) = \frac{1}{\left(k_X^2 + k_Y^2\right)^2 + i\omega}\ .
\label{eq:Greens_fourier}
\end{equation}
We perform the inverse Fourier transform on Eq.~\eqref{eq:Greens_fourier}
to get the Green's function in integral form

\begin{equation}
G(X,Y,T) = \frac{\Theta(T)}{(2\pi)^2}\int \textrm{d}k_X\textrm{d}k_Y\, \,\textrm{e}^{-T(k_X^2+k_Y^2)^2}\,\textrm{e}^{i(k_XX+k_YY)}
\label{eq:greens_integral}
\end{equation}
which can be further expressed (using e.g. Mathematica) through a Meijer's function:

\begin{equation}
G(X,Y,T) = \frac{2\Theta(T)}{\pi^{5/2}(X^2+Y^2)}\textrm{MeijerG}_{6,0}^{0,4}\left[\left\{\left\{0, \frac{1}{4}, \frac{1}{2}, \frac{3}{4}\right\}, \left\{\frac{1}{4}, \frac{3}{4}\right\}\right\}, \frac{2048^2T^2}{(X^2+Y^2)^4} \right].
\label{eq:greens_carthesian}
\end{equation}
Introducing the similarity variable $\xi = \sqrt{X^2+Y^2}/T^{1/4}=RT^{-1/4}$ we can rewrite the Green's function as
\begin{equation}
\label{ssGreen}
G(\xi, T) = \frac{\Theta(T)}{T^{1/2}}f(\xi)
\end{equation}
with

\begin{equation}
f(\xi) = \frac{2}{\pi^{5/2}\xi^2}\textrm{MeijerG}_{6,0}^{0,4}\left[\left\{\left\{0, \frac{1}{4}, \frac{1}{2}, \frac{3}{4}\right\}, \left\{\frac{1}{4}, \frac{3}{4}\right\}\right\}, \frac{2048^2}{\xi^8} \right]
\label{eq:attractor}
\end{equation}
and $\lim_{\xi\to 0} f \approx 0.011224$.

To verify the validity of the rescaled Green's function obtained in Eq.~\eqref{eq:attractor}, we perform a finite-element numerical integration (FENI) of Eq.~\eqref{eq:gtfeadim}, with $\beta=0$ and a Dirac function as initial condition\footnote{For differential operators such as $\mathcal{L}$, and for $T>0$, the Green's function is also the solution of Eq.~\eqref{eq:greens_definition} without right-hand-side term but with an initial profile $G(X,Y,0)=\delta(X,Y)$.}. In Fig.~\ref{fig:profiles}a, we plot the two normalized solutions and we see that the match is perfect, demonstrating the validity of Eq.~\eqref{eq:attractor}. The FENI at several times $T$ (see inset) collapse onto a single curve when rescaling as in Eq.~\eqref{ssGreen}, showing the inherent self-similarity of the glassy levelling process.

The Green's function expressed here is the 3D-axisymmetric equivalent of the 2D one studied in~\cite{benzaquen2013intermediate}, with the normalized Eq.~\eqref{eq:attractor} acting as a universal attractor. This Green's function can now be used to calculate the film profile at any time, by solving Eq.~\eqref{eq:convolution}.

\section*{General axisymmetric solution}

\begin{figure}
\centering
\includegraphics[width=0.48\linewidth]{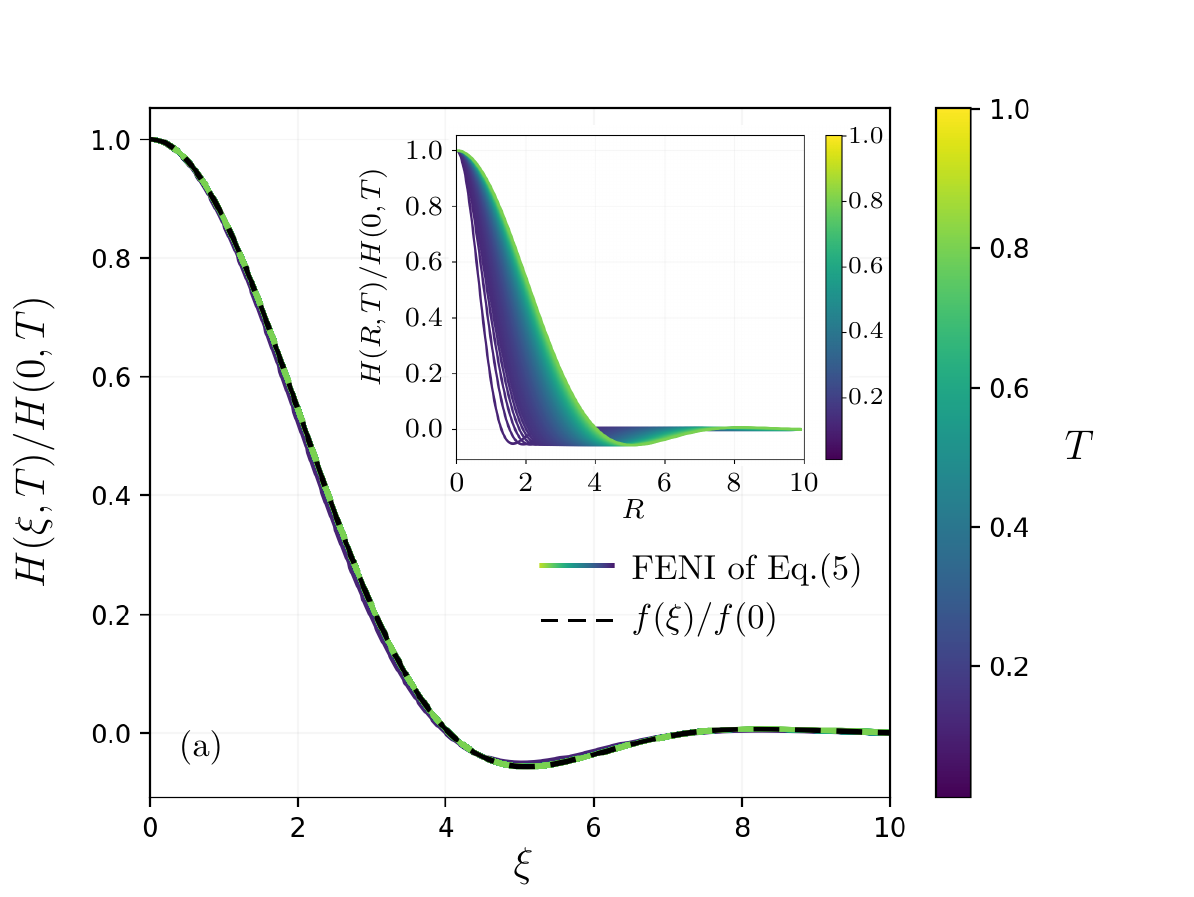}
\includegraphics[width=0.48\linewidth]{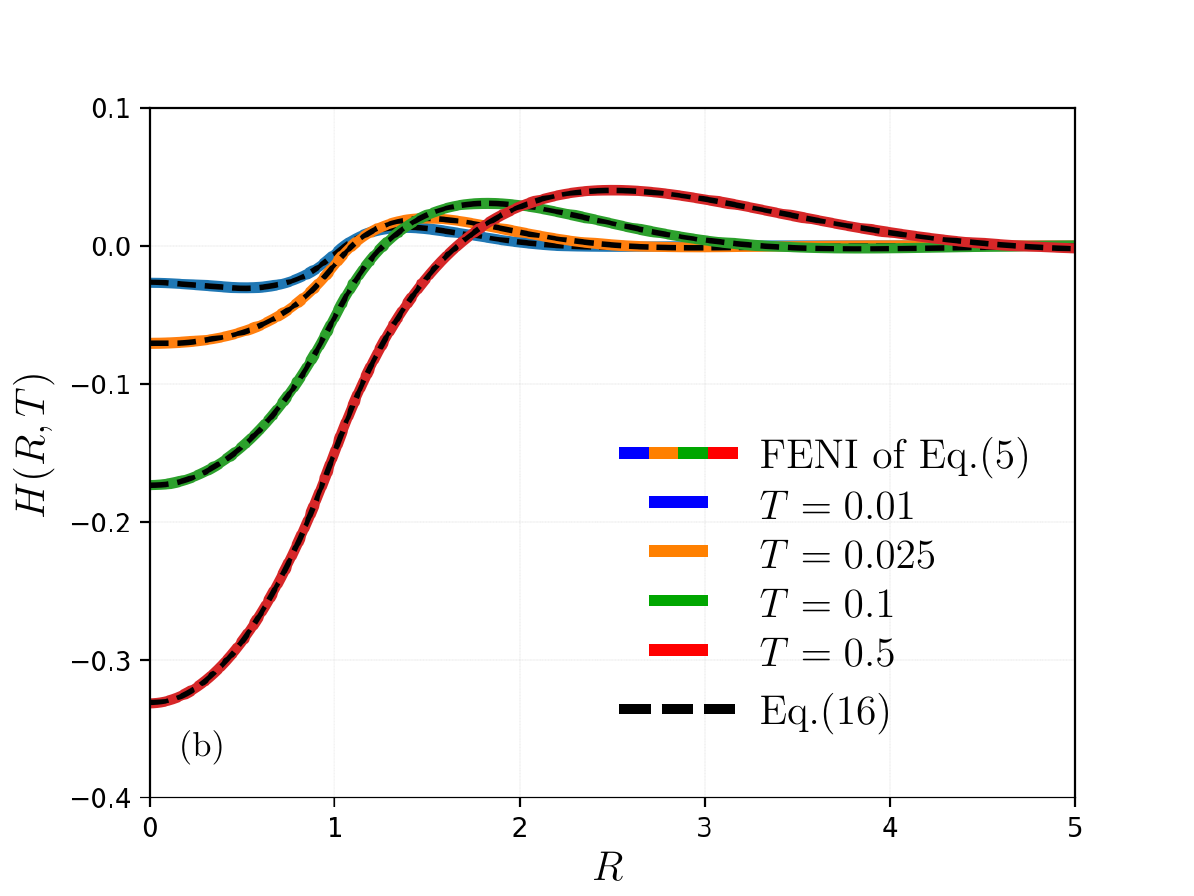}
\caption{(a) Normalized surface profile of a levelling glassy thin film with no bubble, but with a Dirac initial surface perturbation. The dashed line is the normalized Green's function of Eq.~\eqref{eq:attractor}. The solid colored lines represent the normalized finite-element numerical integration (FENI) of Eq.~\eqref{eq:gtfeadim} with $\beta=0$ (no bubble), where the dimensionless times $T$ are given by the color bar.  The inset shows the normalized FENI profiles as functions of the radial coordinate $R$, at different times, while the main figure shows the same data plotted as a function of the similarity variable $\xi=RT^{-1/4}$. (b) Evolution of the surface profile of a glassy thin film induced by the presence of a bubble. The solid lines are the FENI of Eq.~\eqref{eq:gtfeadim} with $\beta=1$, at different times $T$ as indicated, while the dashed lines are the numerical estimates of Eq.~\eqref{eq:solution}.}
\label{fig:profiles}
\end{figure}

As the specific bubble-induced external forcing studied here is axisymmetric, we return to Eq.~\eqref{eq:greens_integral} and perform a change of variables towards polar coordinates: $X=R\cos(\theta)$, $Y=R\sin(\theta)$, $k_X=\rho\cos(\psi)$ and $k_Y=\rho\sin(\psi)$, with $\theta$ the angular coordinate, and $\rho$ and $\psi$ the radial and angular coordinates in Fourier space. We obtain

\begin{equation}
\begin{split}
G(R,T) &= \frac{\Theta(T)}{(2\pi)^2}\int \textrm{d}\rho\, \rho \,\textrm{e}^{-\rho^4T}\int \textrm{d}\psi \,\textrm{e}^{i\rho R \cos(\psi - \theta)} \\
       &= \frac{\Theta(T)}{2\pi} \int \textrm{d}\rho\, \rho \,\textrm{e}^{-\rho^4T}J_0(\rho R)
\end{split}
\end{equation}
where $J_0$ is the zeroth-order Bessel function. By performing a similar change of variables in Eq.~\eqref{eq:convolution} and subsequently inserting  the previous axisymmetric Green's function, we obtain the general film profile in integral form

\begin{equation}
\begin{split}
H(R, T) &= \beta \int \textrm{d}T'\textrm{d}\theta \textrm{d}R'\,  G\left(\sqrt{R^2 + R'^2 - 2RR'\cos(\theta)}, T-T' \right) \Theta(T')\partial_{R'}\left[R'\partial_{R'}\Theta(1-R')\right] \\
        &= \frac{\beta\Theta(T)}{2\pi} \int_0^{\infty}\textrm{d}\rho\, \rho \int_0^{2\pi} \textrm{d}\theta\, \int_0^{T} \textrm{d}T'\, \,\textrm{e}^{\rho^4(T'-T)}  \int \textrm{d}R'\, J_0\left(\rho\sqrt{R^2 + R'^2 - 2RR'\cos(\theta)}\right)\partial_{R'}\left[R'\partial_{R'}\Theta(1-R')\right] \\
        &= \frac{\beta\Theta(T)}{2\pi} \int_0^{\infty} \textrm{d}\rho\, \frac{1-\,\textrm{e}^{-\rho^4 T}}{\rho^2}\int_0^{2\pi}\textrm{d}\theta\, \frac{R\cos(\theta)-1}{\sqrt{R^2 + 1 - 2R\cos(\theta)}} J_1\left(\rho\sqrt{R^2 + 1 - 2R\cos(\theta)}\right)\ .
\end{split}
\label{eq:solution}
\end{equation}
As expected from the linearity of the glassy thin film equation, the response is proportional to $\beta$. In particular the profile perturbation $H$ vanishes in the absence of any bubble ($\beta=0$). Besides, as $T$ goes to zero by positive values, we recover a vanishing surface perturbation $H$. We can now evaluate Eq.~\eqref{eq:solution} numerically at different times. In Fig.~\ref{fig:profiles}b, such an evaluation is compared  to the FENI of Eq.~\eqref{eq:gtfeadim} with $\beta=1$, at four different times. The agreement is perfect, thus validating the results and methods.

\section*{Depression at the bubble center ($R=0$)}

\begin{figure}
\centering
\includegraphics[width=0.48\linewidth]{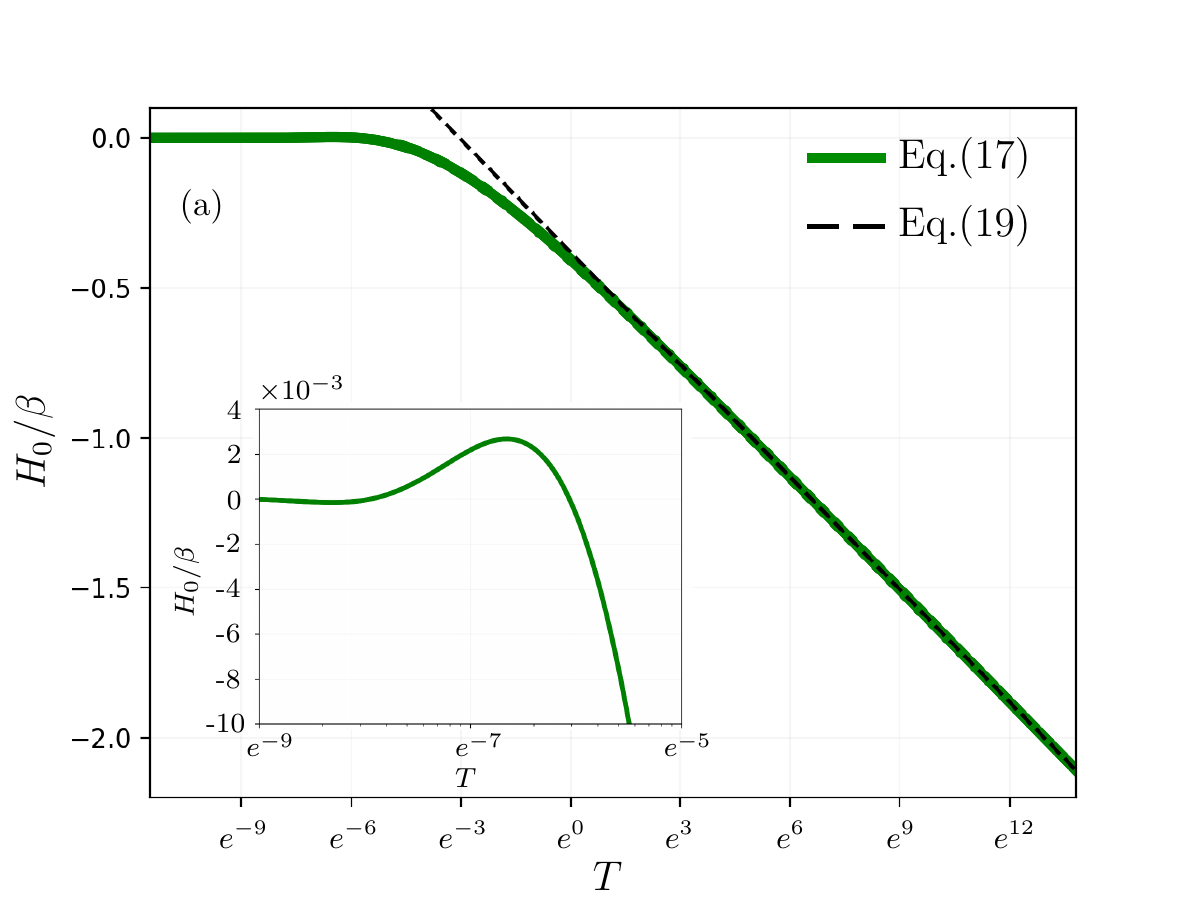}
\includegraphics[width=0.48\linewidth]{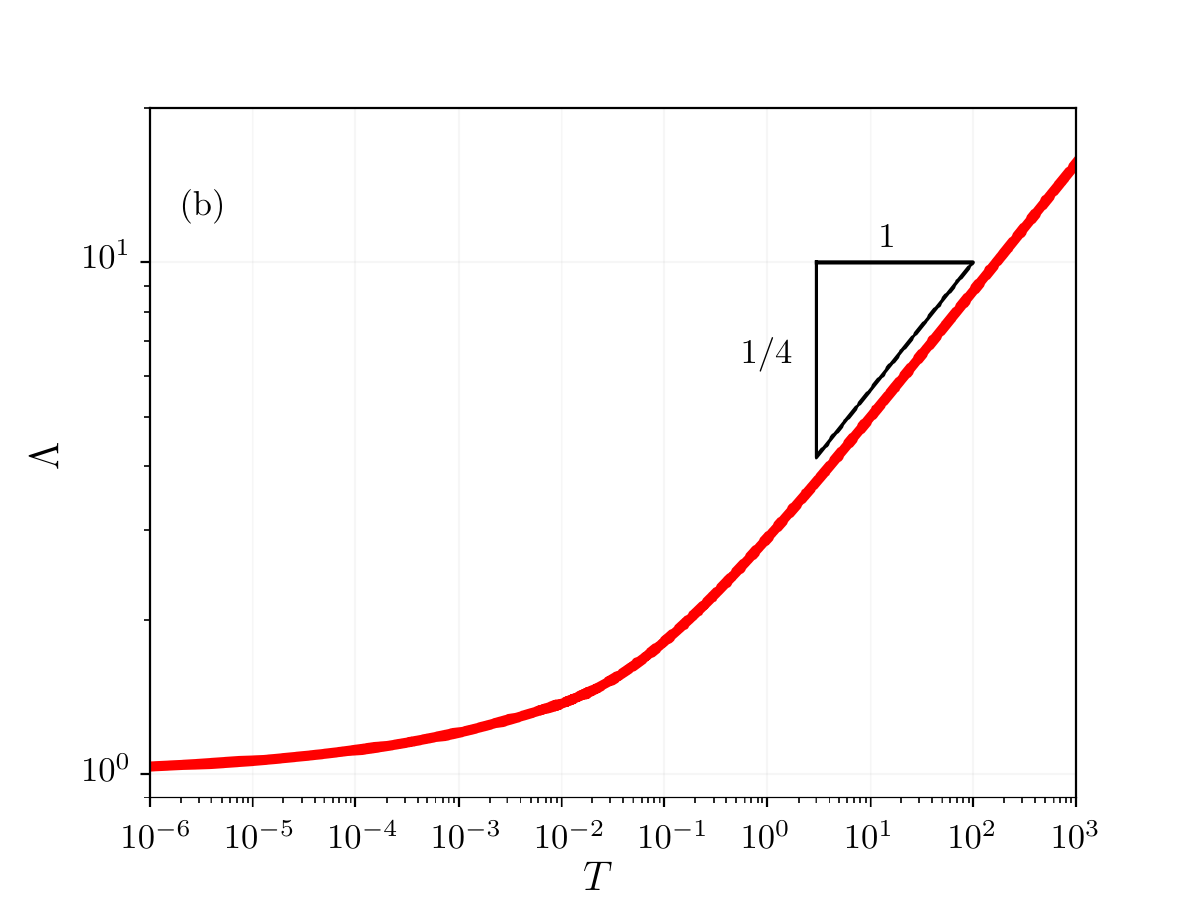}
\caption{(a) Normalized central height of the bubble-induced perturbation of the film profile as a function of time, computed from Eq.~\eqref{central}. The dashed line indicates the asymptotic expression of Eq.~\eqref{asym}, with $C=0.04746$. (b) Half width of the bubble-induced perturbation as a function of time, defined as the radial coordinate of the maximum of $H(R,T)$ (see Eq.~\eqref{eq:solution} and Fig.~\ref{fig:profiles}b). The slope triangle indicates a $1/4$ power-law exponent.}
\label{fig:height_and_radius}
\end{figure}
At $R=0$, Eq.~\eqref{eq:solution} reduces to a single integral, defining the central depth $-H_0(T)=-H(0,T)$ through

\begin{equation}
\label{central}
H_0(T) = \beta \Theta(T)\int_0^{\infty} \textrm{d}\rho\, J_1(\rho) \frac{\,\textrm{e}^{-\rho^4 T}-1}{\rho^2}\ .
\end{equation}
We numerically evaluate the latter function and plot the normalized result in Fig.~\ref{fig:height_and_radius}a.
There are two interesting characteristics to observe. First, we see in the inset that there is an increase in the central height of the film at early times. This is attributed to the sharp spatial pressure profile, which generates traveling surface waves in both the forward and backward radial directions.
Secondly, beyond $T\approx 1$, $H_0(T)$ reaches a logarithmic asymptotic behaviour in time. Indeed, at large $T$, and apart from additive constants, Eq.~\eqref{central} can be well approximated by
\begin{equation}
\begin{split}
H_0(T) &\simeq -\beta \int_{T^{-1/4}}^{\infty} \textrm{d}\rho\,  \frac{J_1(\rho)}{\rho^2}\\
&\simeq-\frac{\beta}2 \int_{T^{-1/4}} \frac{\textrm{d}\rho}{\rho}
\end{split}
\end{equation}
where we invoked the first-order Taylor expansion of $J_1$ near the origin, and only considered the lower bound of the integral as it drives the divergence in time. By including the integral constant $C$, one gets

\begin{equation}
H_0(T)\simeq\frac{\beta}8\ln\left(\frac CT\right)\ .
\label{asym}
\end{equation}
The latter expression matches well the long-term behaviour in Fig.~\ref{fig:height_and_radius}a, with $C=0.04746$.

In dimensional units, with $h_0(t)=h(0,t)=h_{\textrm{m}}H_0(T)$, one obtains the asymptotic expression
\begin{equation}
h_0(t) \simeq \frac{r_{\textrm{b}}\gamma_{\textrm{LV}}}{4\gamma}\ln\left(\frac{3C\mu r_{\textrm{b}}^4}{\gamma h_{\textrm{m}}^3t}\right)\ .
\end{equation}
An important consequence of this theoretical prediction is that any glassy film of finite thickness $h_{\textrm{i}}$ will eventually dewet if exposed to surface nanobubbles during a given finite time. For small enough bubbles and/or thick enough films, the kinetics is essentially determined by the sole asymptotic regime, and the dewetting criterion $h_0(t_{\textrm{d}})=-h_{\textrm{i}}$ leads to the following prediction for the dewetting time

\begin{equation}
t_{\textrm{d}}=\frac{3C\mu r_{\textrm{b}}^4}{\gamma h_{\textrm{m}}^3}\exp\left(\frac{4\gamma h_{\textrm{i}}}{\gamma_{\textrm{LV}}r_{\textrm{b}}}\right)\ .
\end{equation}
Hence, the dewetting time $t_{\textrm{d}}$ grows exponentially with the ratio between film thickness and bubble size, and is proportional to the inverse mobility of the surface mobile layer, which offers a way to infer the latter fundamental quantity.

\subsection*{Half width}
We define the half-width $\Lambda(T)$, as the radial coordinate $R=\Lambda(T)$ at which $H(R,T)$ is maximum (see Fig.~\ref{fig:profiles}b).
We compute it numerically from Eq.~\eqref{eq:solution} for a large set of times, and plot the results in Fig.~\ref{fig:height_and_radius}b.
Beyond $T\approx 1$, we observe a $R(T)\sim T^{1/4}$ power-law, naturally emerging from the inherent self-similarity of the glassy thin film equation (see Fig.~\ref{fig:profiles}a). Interestingly, this behaviour holds at large perturbations, until dewetting, which is a direct signature of the localized surface mobility in glasses, in sharp contrast with the Tanner-like regime in liquid films~\cite{cormier2012beyond}. We further stress that the lateral power-law spreading is faster than the vertical logarithmic decay discussed above, ensuring the validity of the small-slope approximation at late times.

\subsection*{Surface energy}

\begin{figure}
\centering
\includegraphics{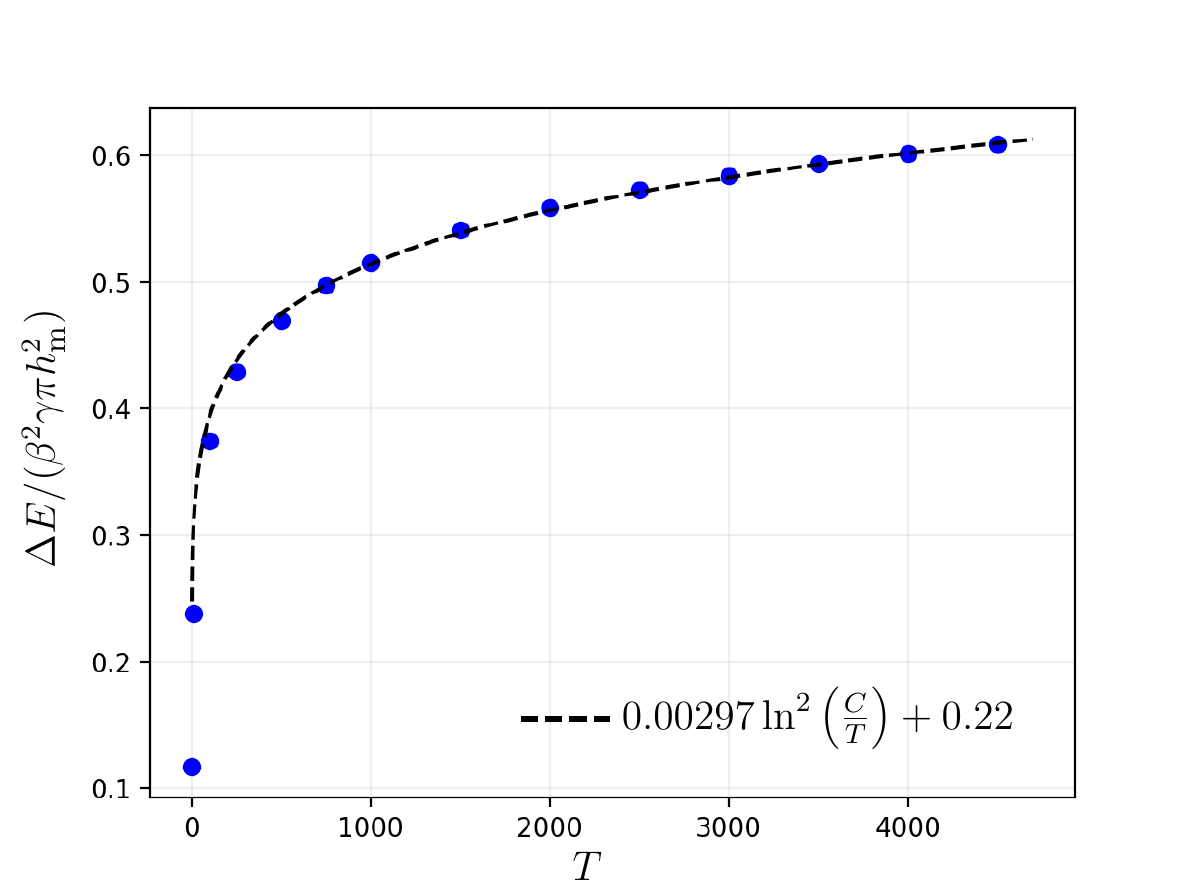}
\caption{Normalized excess surface energy of the film as a function of time, as computed from Eq.~\eqref{excess} and the solution in Eq.~\eqref{eq:solution}, for $\beta=1$. The dashed line is a best fit to Eq.~\eqref{logsquared} with a numerical prefactor $0.00297$ and an offset 0.22.}
\label{fig:area}
\end{figure}
As the surface is deformed by the bubble, its area increases, leading to an increase in the surface energy of the film. For small slopes, the excess surface energy with respect to the flat initial configuration reads

\begin{equation}
\begin{split}
\Delta E &\simeq \gamma\pi\int \textrm{d}r\, r\left(\frac{\partial h}{\partial r}\right)^2\\
&\simeq \gamma\pi h_{\textrm{m}}^2\int \textrm{d}R\, R\left(\frac{\partial H}{\partial R}\right)^2\ .
\end{split}
\label{excess}
\end{equation}
From scaling arguments, we expect $\Delta E/(\gamma\pi h_{\textrm{m}}^2)\sim H_0(T)^2$. Invoking Eq.~\eqref{asym}, this leads to the asymptotic scaling

\begin{equation}
\frac{\Delta E}{\gamma\pi h_{\textrm{m}}^2\beta^2}\sim \ln^2\left(\frac CT\right)\ .
\label{logsquared}
\end{equation}
In Fig.~\ref{fig:area}, we plot the dimensionless excess surface energy $\Delta E/(\beta^2\gamma\pi h_{\textrm{m}}^2)$ as a function of time, as computed from Eq.~\eqref{excess} and the solution in Eq.~\eqref{eq:solution} for $\beta=1$. The result is well fitted by Eq.~\eqref{logsquared} with a numerical prefactor $0.00297$ and an offset $0.22$.

\section*{Non-linear effects}

\begin{figure}
\centering
\includegraphics{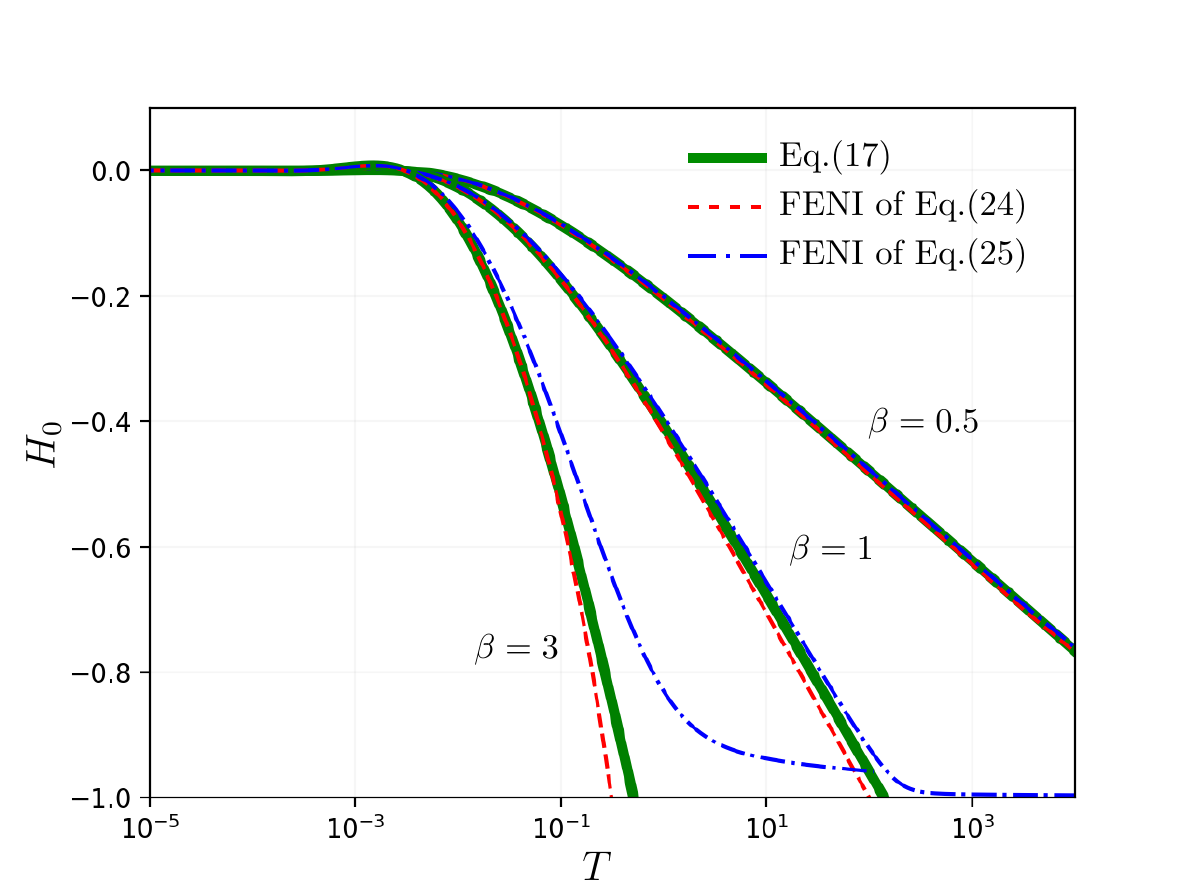}
\caption{Central height of the perturbation of the film profile as a function of time, as obtained from the solutions of the GTFE (Eq.~\eqref{central}), MTFE (FENI of Eq.~\eqref{eq:gtfeadim_mod} with $(h_{\textrm{m}}/r_{\textrm{b}})^2$=0.1) and TFE (FENI of (Eq.~\eqref{eq:gtfeadim_nonlin}), for three different $\beta$ as indicated.}
\label{fig:non-linear}
\end{figure}

In this final section, we investigate the role of non-linearities, resulting either from non-linear curvature effects, or from the film profile in ultrathin films.

First, to investigate non-linear curvature effects, we return to Eq.~\eqref{eq:pressure} and avoid the small-slope approximation in the curvature by including the full norm $|\mathbf{n}| = \sqrt{1 + (\partial h/\partial r)^2}$ of the normal vector. This does not impact the validity of the lubrication approximation provided that the typical scale separation $(h_{\textrm{m}}/r_{\textrm{b}})\ll1$ is maintained, which we ensure in the following numerical tests. When inserting the modified pressure term into Eq.~\eqref{eq:gtfe}, we obtain the modified GTFE (MGTFE) with an external driving pressure, which reads in dimensionless form

\begin{equation}
\frac{\partial H(R,T)}{\partial T} + \frac{1}{R} \frac{\partial}{\partial R}\left\{R\frac{\partial}{\partial R}\left[\frac{1}{R}\frac{\partial}{\partial R}\left(R\frac{\partial_RH(R,T)}{\sqrt{1+(h_{\textrm{m}}/r_{\textrm{b}})^2(\partial_R H)^2}}\right) - \beta\Theta(T)\Theta(1-R) \right] \right\} = 0\ .
\label{eq:gtfeadim_mod}
\end{equation}

Secondly, when the total thickness of the film becomes similar to, or smaller than, the thickness of the mobile layer, the whole film flows and Eq.~\eqref{eq:gtfeadim} should be replaced by the non-linear capillary-driven thin film equation (TFE)~\cite{cormier2012beyond} with external driving, which reads in dimensionless form

\begin{equation}
\frac{\partial H(R,T)}{\partial T} + \frac{1}{R} \frac{\partial}{\partial R}\left\{H^3R\frac{\partial}{\partial R}\left[\frac{1}{R}\frac{\partial}{\partial R}\left(R\frac{\partial}{\partial R} H(R,T)\right) - \beta\Theta(T)\Theta(1-R) \right] \right\} = 0\ .
\label{eq:gtfeadim_nonlin}
\end{equation}

In Fig.~\ref{fig:non-linear}, we plot the central magnitudes $H_0(T)$ as functions of time, as obtained from the solutions of the GTFE, MTFE and TFE, for three different $\beta$. For $\beta=0.5$ and below, there is no noticeable difference between the three solutions within the considered temporal range. However, as $\beta$ increases, so do the differences. The MGTFE, and thus the non-linear curvature, appear to accelerate slightly the dewetting process in the considered $\beta$ range. In contrast, the TFE, and thus the profile non-linearities, impede the dynamics near dewetting. This slowing down is expected since the size of the flowing layer vanishes in the TFE description, acting as a regularization mechanism to the dewetting process discussed above.

All together, these effects stress the importance of non-linearities at large values of $\beta$,
and/or when the film thickness approaches zero. In the latter case, we must however mention that close to film rupture other effects will come into play and are expected to dominate the dynamics, such as
van der Waal forces~\cite{ida1996thin} and the altered polymer entanglement density~\cite{tsui2001effects, Si2005,hoy2006strain}.

\section*{Conclusion}

We reported on the theoretical treatment of the nanobubble-induced instability of a thin glassy polymer film immersed in water, due to the existence of a surface mobile layer. By using the Green's function formalism, we obtained a semi-analytical solution of the axisymmetric glassy thin film equation with an external source term describing the Laplace pressure of the bubble. We further characterized the solution, by extracting key dynamical quantities such as the central depth, half width, and excess surface energy of the film. In particular, we demonstrated the existence of an asymptotic logarithmic temporal increase of the central depth of the perturbation, leading to a dewetting scenario at finite time. The dewetting time was obtained analytically, growing exponentially with the ratio between film thickness and bubble size, and being proportional to the inverse mobility of the surface mobile layer. Finally, we investigated the corrections to this scenario induced by curvature and profile non-linearities. Our predictions might be useful for determining the fundamental mobility of glassy materials, and may have practical implications on the stability, patterning and creation of nanoporous membranes.

\section*{Acknowledgments}

Y. W.and S. R. appreciate financial support from the National Natural
Science Foundation of China (Grants No. 51775028 and
No. 52075029)

\bibliographystyle{ieeetr}
\bibliography{nanobubbles}

\end{document}